\documentclass[12pt,tightenlines]{revtex4-1}   %% LaTeX 2e (preferred)
\usepackage{graphicx}
\usepackage{amsmath}
\usepackage{hyperref}
\hypersetup{
  pdfauthor={Harald G. L. Schwefel},
  pdftitle={Directional Emission of Dielectric Disks with a Finite Scatterer in the THz Regime},
  pdfsubject={Physics, Optics, Whispering Gallery Mode Resonators},
  urlcolor=blue,
}

\usepackage[english]{babel}

\newcommand{\degree} {$^\circ$}

\newcommand{\be}{\begin{equation}}
\newcommand{\ee}{\end{equation}}

\newcommand{\bild}[5]{\begin{figure}[#3]
                        \centering
                        \includegraphics*[width=#2]{#1}%,height=#2
                        \sl\caption{#4}
                        \label{#5}
                        \end{figure}}

\newcommand{\Figref}[1]{Fig.~\ref{#1}}%\usepackage{arev}

\begin{document}

\title{Directional Emission of Dielectric Disks with a Finite Scatterer\\ in the THz Regime}
\author{S. Preu,$^{1}$ S. I. Schmid,$^{2,3}$ F. Sedlmeir,$^{4,5,6}$\\ J. Evers,$^2$ H. G. L. Schwefel$^{4,5}$}

\affiliation{$^1$Chair for Applied Physics, University of Erlangen-N\"urnberg,
D-91058 Erlangen, Germany}
\affiliation{$^2$Max Planck Institute for Nuclear Physics, Saupfercheckweg 1, 69117 Heidelberg, Germany}
\affiliation{$^3$present address: Beijing Computational Science Research Center, No. 3 He-Qing Road, Hai-Dian District, Beijing, 100084 China}
\affiliation{$^4$Institute for Optics,
Information and Photonics, University of Erlangen-N\"urnberg,
D-91058 Erlangen, Germany}
\affiliation{$^5$Max Planck Institute for the Science of Light,
D-91058 Erlangen, Germany}
\affiliation{$^6$SAOT, School in Andvanced Optical Technologies, Paul-Gordan-Str.\ 6,
D-91052 Erlangen, Germany}
\email{Sascha.Preu@physik.uni-erlangen.de}
\author{\today}

\begin{abstract}
In the Terahertz (THz) domain, we investigate both numerically and experimentally the directional emission of whispering gallery mode resonators that are perturbed by a small scatterer in the vicinity of the resonators rim. We determine quality factor degradation, the modal structure and the emission direction for various geometries. We find that scatterers do allow for directional emission without destroying the resonator's quality factor. This finding allows for new geometries and outcoupling scenarios for active whispering gallery mode structures such as quantum cascade lasers and passive resonators such as evanescent sensors. The experimental results agree well with finite difference time domain simulations. 
\end{abstract}
\maketitle

\section{Introduction}
 The THz frequency range, defined as the frequency range from 100 GHz to 10 THz, corresponding to a wavelength between 3 mm and 30 $\mu$m offers unique advantages for studying whispering gallery mode (WGM) resonators. On the one hand, much larger resonators and feature sizes have to be used compared to the optical domain. In the lower THz range around 100-300 GHz, the wavelength is roughly a factor of 3000 larger than in the visible domain. However, optical and physical effects remain the same due to the scalability of electromagnetism. It is therefore much easier to characterize the near and far field \cite{preu_coupled_2008}. On the other hand, materials are less advanced than in the optical domain, resulting in losses and therefore much lower quality factors. Furthermore, there is a lack of powerful, but at the same time, tunable continuous-wave sources. We developed a THz system based on n-i-pn-i-p superlattice photomixers \cite{preu_efficient_2007,ReviewSP} that allows for 
characterizing WGM resonators between 60 GHz and several 100 GHz \cite{preu_coupled_2008}. 
 
 In this paper, we study the emission characteristics of WGM resonators in the lower THz range. For most applications of WGM resonators, not only the modal distribution within the resonator, but also knowledge on the radiative losses due to the shape and structure of the resonator are extremely important. Several detection schemes employ the evanescent field surrounding a WGM resonator for sensing. Gas, fluids, or particles surrounding the resonator influence the evanescent field by either shifting the effective refractive index or by absorption \cite{vollmer_whispering-gallery-mode_2008,loock_absorption_2010}. Consequently, this alters the modal structure of the resonator, resulting in mode shifts and altered quality ($Q$) factors. For WGM lasers, particularly THz quantum cascade lasers \cite{kohler_terahertz_2002}, the directional out-coupled emission is paramount. Therefore, the resonator shape often deviates from the circular geometry. Such examples include chaotic micro resonators \cite{gmachl_high-power_1998} where the light is confined on a stable periodic lightray-orbit or by the short term dynamics of the chaotic light ray-dynamics inside the resonator which 
channels a direct emission port \cite{schwefel_dramatic_2004,song_channeling_2012}. 
We present another way to achieve directional emission: In this paper, we study directional emission from a WGM resonator with a finite scatterer both experimentally and theoretically in the THz frequency range. Theoretical calculations by Wiersig et al.\ \cite{wiersig_unidirectional_2006} already described efficient outcoupling of a high $Q$ mode via interaction with a low $Q$ mode without ruining the $Q$ factor. Here, however, we address the mode directly by perturbing the mode with the scatterer. We will show that the perturbation may be weak enough to maintain a considerably high $Q$ factor while the power is coupled out directionally. We will use passive resonators that are coupled to a waveguide. The paper is structured as follows: First, we measure the influence of the waveguide on the modal structure and near field of the resonator by a near field probe. Second, we experimentally determine the influence of the position of a hole acting as scatterer on the $Q$ factor of the disk. Third, we will explain the numerical algorithm that was used to theoretically determine the radiation pattern of a disk with hole. Last, we will show experimental results on the outcoupling performance and compare the results to theory.

%{\tt  -- this is way too negative and should be in the introduction, it should also have a section telling people why THz is so great!!! --}

\section{Experimental Setup}
 We use a n-i-pn-i-p superlattice photomixer \cite{ preu_tunable_2011} as THz source that has been developed in house. The measurement setup is illustrated in Fig.~\ref{FigSetup} a) and b). 
The source is mounted on a silicon lens for pre-collimation and efficient out-coupling of the THz beam. The beam is then collimated with a parabolic mirror (PM1).  A second parabolic mirror (PM2) focuses the THz beam on a feed horn (FH) with a rectangular Teflon waveguide (dimensions 1 mm x 1.5 mm) mounted inside the horn. The feed horn is used to increase the coupling efficiency. The Teflon waveguide delivers the THz power to the WGM resonator. A Golay cell detector (D) is used to monitor the transmission through the waveguide. The WGM resonator is situated in the center of a rotation stage. The Golay cell can be mounted on the rotation stage to scan the far field emission from the resonator. A horizontal cylindrical (HL) lens collects THz power emitted perpendicular to the detection plane. A second, vertical cylindrical lens (DL) is mounted in front of the Golay cell to collect power within a 10\degree-15\degree angle to improve the signal to noise ratio. The scan range was limited by the Teflon waveguide and 
the size of the Golay cell and the lenses to about 140$^\circ$. Alternatively, the Golay cell detector can also be attached to a probe waveguide that touches the resonator in order to scan the near field. In order to reduce the necessary integration times of the Golay cell novel field-effect transistors could also be used \cite{preu_improved_2012,preu_detection_2012}
\bild{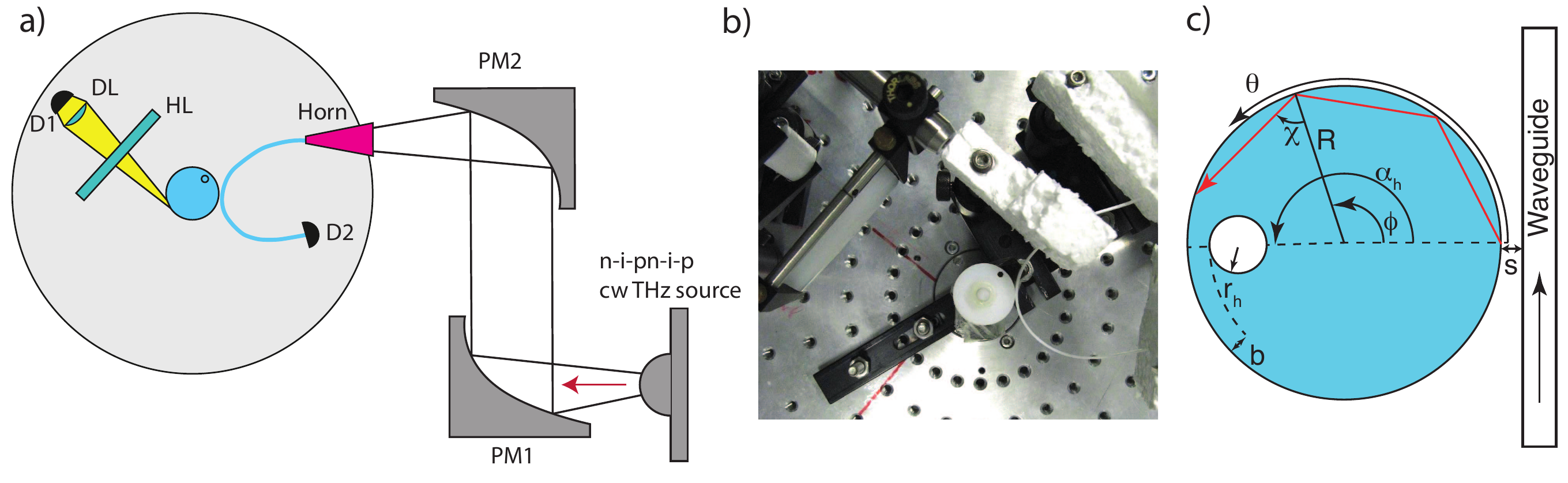}{\linewidth}{t}{ Schematic setup. The resonator is mounted at a fixed position in the center of a rotation stage. The detector (Golay cell) scans the angular far field pattern of the disk. Two cylindrical lenses (horizontal lens, HD, and detector lens, DL) improve the signal to noise ratio by focusing the emitted power from the resonator on the detector. For most measurements, the lenses collect radiation within a 10\degree angle. Alternatively, the lenses can be replaced with a probe waveguide that touches the boundary of the disk to scan the near field. b) Photograph of the resonator in the center of the ration mount. c) Definition of variables used throughout the paper. Angles are defined with respect to the coupling position, the angle of incidence is defined as positive in the direction of the light. The thickness of the boundary between hole and disk circumference is labelled as $b$. The hole (radius $r_h$) position relative to the coupling position is defined as $\alpha_h$. Parameters for the numerical calculation, such as the incidence angle of a light ray on the surface, $\chi$, and the far field emission angle $\theta$ are also included.}{FigSetup}

The frequency resolution  (120 MHz) of the system is sufficient to measure $Q$ factors up to 1200 in the frequency range from $100 - 300$ GHz. Due to the lack of ultra-low-loss materials at THz frequencies, this resolution is sufficient since the ideal loss-limited Q factors are in the range of a few 1000. As an example, the imaginary part of the dielectric constant of polyethylene and Teflon \cite{microtech} is in the range of  $\varepsilon''>1.8\times 10^{-3}$, resulting in an absorption-limited $Q$ factor of $<1100$, even if no radiative losses are present. For the investigated disks and frequency ranges, however, the measured $Q$ factor was always smaller than 1200, limited by radiative losses.

\begin{figure}[tbp]
\centering\includegraphics[width=\linewidth]{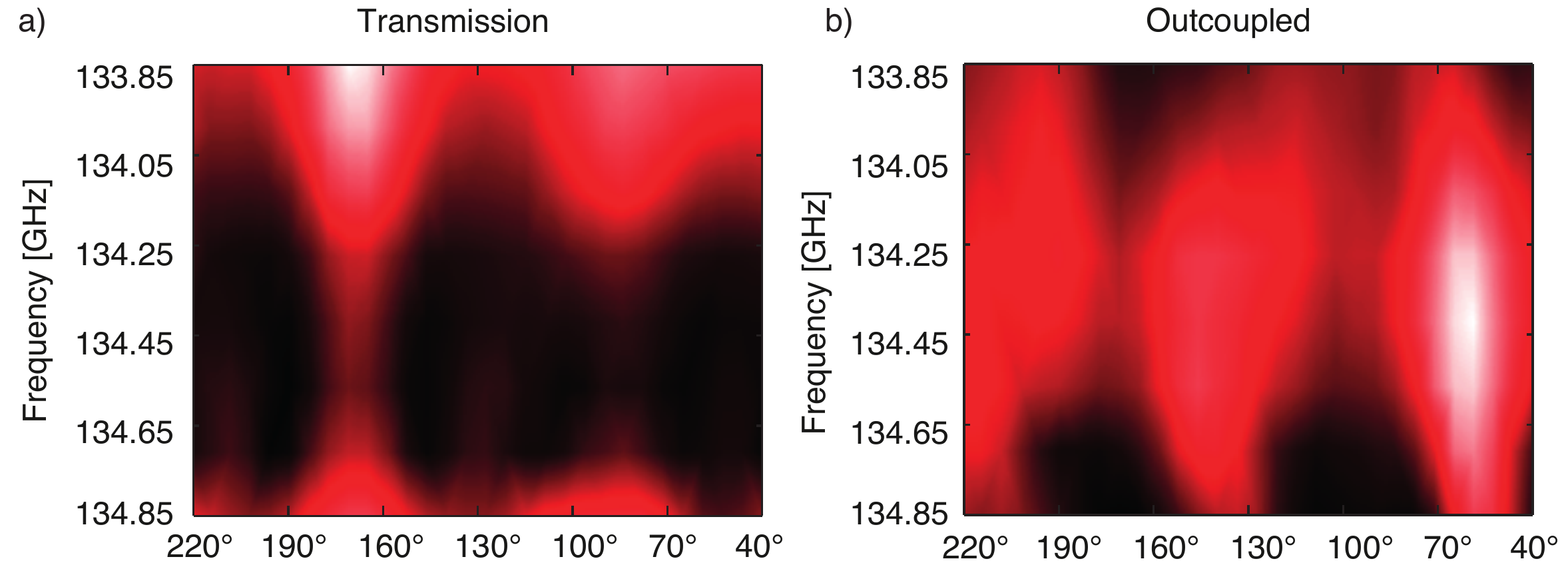}
\caption{The evanescent near field of a polyethylene WGM resonator (diameter 25 mm, in the following referenced as ``large resonator") without hole is probed with a waveguide. a) Transmission through the feed waveguide while the near field probe is scanned along the circumference of the resonator. b) Outcoupled power through the near field probe.}\label{fig:2Dimage}
\end{figure}

In a first step, we investigate the perturbation of the coupler on the modal structure of the resonator without scatterer. An unperturbed circular resonator should show a position-independent modal field strength. We resolved the local field strength by a near field tapered probe waveguide that weakly couples to the disk. In Fig.~\ref{fig:2Dimage} a) we show the transmitted power through the feed waveguide and in Fig.~\ref{fig:2Dimage} b) the outcoupled power through the probe while scanning both frequency across a resonance and the position of the probe. As expected, the transmitted power through the feed waveguide is high, when little power is coupled out through the probe waveguide. However, we also see a strong, periodic probe position dependence of the power in both sub-figures with a five fold symmetry. This number is much too small to be the modal number ($\sim 50$) for a resonator with a diameter of 25 mm at a wavelength of 2.24 mm. Already the coupling to the feed waveguide perturbs the modal structure of the resonator considerably. The orbit of the mode is not circular any more, there exist positions where the mode is close to the surface, allowing for enhanced outcoupling, and further away, showing local minima in Fig.~\ref{fig:2Dimage} b).

\begin{figure}[tbp]
\centering\includegraphics[width=\linewidth]{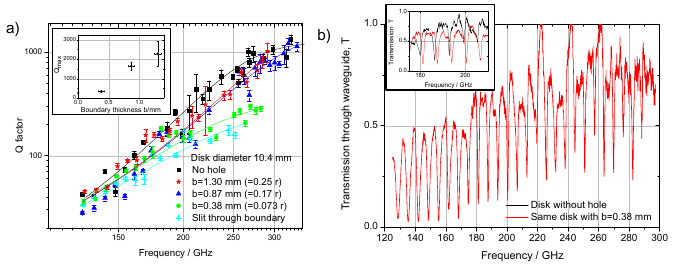}
\caption{a) $Q$ factors of several disks without and with a hole with a boundary thickness of $b$ (see \Figref{FigSetup}). The closer the hole is drilled to the resonator circumference, the stronger are the perturbations of the mode and the stronger the reduction of the $Q$ factor. The inset shows the extrapolated loss-limited $Q$ factor in the limit of large frequencies. b) Modal structure of the disk with $b=0.38$ mm before and after the hole was drilled. Despite a strong shift in the resonance frequency, the $Q$ factor was only slightly altered.}\label{figdiameterQ}
\end{figure}

As a next step, we investigated the influence of the hole position on the quality factor of the resonances with disks with a diameter of $d=10.5$ mm at wavelengths between 2.3 mm and 0.91 mm (0.13-0.33 THz). For holes very far away from the circumference, the outermost radial modes that are excited by the waveguide, are not affected by the hole as their mode volume does not overlap with it. The $Q$ factors are similar to that of a resonator without hole. When holes are drilled closer to the circumference, the overall $Q$-factor decreases as shown in Fig.~\ref{figdiameterQ} and the mode position shifts because the mode gets perturbed by the hole. The effect becomes more pronounced if the boundary thickness $b$ (see \Figref{FigSetup} c)) becomes comparable to or smaller than the wavelength. The loss-limited quality factor can be estimated using an empiric fit of the form

\be
Q(\nu)\approx \frac{Q_{loss}Q_0\exp{(\gamma \nu)}}{Q_{loss}+Q_0\exp{(\gamma \nu)}},
\ee
where $\gamma$ and $Q_0$ are fitting factors describing the increase of $Q$ with increasing frequency. The formula takes the roughly exponential increase of the quality factor of an unperturbed resonator with increasing frequency into account, that is finally limited by losses to a maximum value of $Q_{loss}$. $Q_0\ll Q_{loss}$ is the extrapolated quality factor at low frequencies. The inset of \Figref{figdiameterQ} shows the extrapolated values for $Q_{loss}$. The disks with $b=0.38$ mm and the disk with the slit are already loss-limited within the measurement range. %Since the setup does not permit measurements above $Q=1200$, the values for the weakly perturbed resonators are only rough estimates.

The $Q$ factors did not degrade drastically due to the perturbation of the hole. This demonstrates that finite scatterers can be used for resonators and laser cavities. In the following, we discuss the theoretical framework for determining the outcoupling performance by the hole.

\section{Theoretical Modeling}
\label{sysma}

We now turn to our theoretical approach to model the experimental data. Such modeling is desirable mainly for two reasons. On the one hand, the model gives insight into the physical mechanism for the directional emission, not least since it provides access to system properties which may be difficult to obtain in the experiment. On the other hand, in order to design the emission pattern of WGM resonators towards specific applications, an accurate theoretical modeling is desirable in order to perform numerical parameter studies. For stationary resonance problems a number of numerical methods exist, such as scattering methods \cite{tureci_modes_2005}, boundary integral methods \cite{wiersig_boundary_2003,tureci_efficient_2007,zou_quick_2011} and multipole methods \cite{schwefel_improved_2009}. These methods however do not allow for studying the effects of the coupled waveguide.

The light propagation is thus simulated by the Finite-Difference Time-Domain (FDTD) Method~\cite{hagness_fdtd_1997,FDTDATaflove,schmid_pathway_2011}. The geometry of waveguide and resonator as well as the material properties are chosen as in the experiment, without fine-tuning of any of the parameters. We however restrict the simulation to a two dimensional rectangular grid in the resonator plane, with grid constant $\Delta x=\Delta y=0.02$~mm. This corresponds to about $\lambda/90$ or $\lambda/110$ for the two wavelengths in the mm range we consider below. 
In two dimensions, Maxwell's equations for the six components of $\vec{E}$ and $\vec{H}$ separate into two disjunct sets of three equations each. As in the experiment, we study the propagation of the EM field in TE$_z$ polarization, such that $H_z$, $E_x$ and $E_y$ need to be considered. We drive the $H_z$ component using a soft source placed at one edge of the waveguide, with a transverse mode profile obtained by solving the corresponding Helmholtz equation. Unwanted reflections from the boundaries of the simulation area are suppressed by Berenger type perfectly matched absorbing boundaries~\cite{berenger1994}.

To predict the emission pattern, we proceed in three steps. First, we excite the system with a temporally short and spectrally broad pulse and monitor the transmission at the waveguide edge opposite to the source. By relating this result to a reference calculation without the resonator, we obtain the modal spectrum of the resonator. We then choose a single resonator mode by matching the numerically obtained mode spectrum to the experimentally observed one. 

In the second step, we excite the system with a monochromatic continuous-wave field at the resonance frequency of the mode identified in the first step, and evolve the system into its stationary state  (modulo the oscillations at the incident field frequency). This evolution can be monitored, e.g., via the time-dependent power flux through the waveguide downstream the resonator. The steady state, where the resonator is fully loaded, is reached on time scales short enough to neglect the initial cavity build-up time.

In the third step, we calculate the emission pattern of the resonator. A principle problem arises from restraining the computation time to a moderate level. The FDTD simulations therefore are constrained to a small region around the resonator ($\sim$ tens of mm) which are small compared to the experimental detector distance ($\sim$~10~cm). Thus it is impractical to include the detector itself into the simulation. But while the detector exclusively monitors the far field radiation component of the emission, the vicinity of the resonator is strongly influenced by near field contributions. Thus, a transformation of the FDTD results into the far field is required \cite{hagness_fdtd_1997,FDTDATaflove,schmid_pathway_2011}.
To this end, we calculate the amplitude and the phase of the electric and magnetic fields on a circle around the resonator center, with radius larger than the resonator radius by a variable offset $\rho$. At each FDTD grid point  $\vec{r}$ on the circle, we fit the FDTD time evolution over few cycles of the incident field  to the function $a\,\sin(\omega t +\phi)$. Here, $\omega$ is the frequency of the incident field; the amplitude $a>0$ and the phase $\phi$ are the fit parameters. 
From the field configuration, we evaluate the time averaged Poynting vector $\vec{S}(\vec{r})$ at each point $\vec{r}$ on the circle. Next, we determine the intersection point $\vec{R}(\vec{r})$ of a ray starting  at $\vec{r}$ in the direction $\vec{S}$ with the detection sphere of radius $R$ around the resonator center. As a result,  the point $\vec{r}$ contributes with magnitude $|\vec{S}|$ to the detection signal at detector position $\vec{R}(\vec{r})$.
By incoherently summing up the contribution of all points on the circle around the resonator we obtain a prediction for the emission pattern as observed by the detector.

We compared the emission patterns evaluated for different offsets $\rho$ from the resonator edge, and found that for small offsets, the results strongly depend on $\rho$. However, already with offsets $\rho$ of few wavelengths of the incident light field, the radiation pattern becomes largely independent of the offset, indicating the suppression of near field components. It should be noted, however, that neither the simulation range nor the detector distance in the experiment are large enough to neglect the finite size of the resonator.

\section{Comparison of Theoretical and Experimental Results}
\label{results}

%\subsection{Comparison to FDTD Results}
%\subsection{Detector Method results}
We investigated two types of resonators for comparison with our theoretical framework. First we consider a polyethylene WGM resonator with $R=5.2$~mm, $r_h=0.51$~mm, $b=0.41$~mm and $\alpha_h=90^\circ$. The relative permittivity is $\varepsilon_{WGM}=2.56$ ($n_{PE}=1.6$) for the resonator and $\varepsilon_h=1$ (air) inside the hole. This resonator is referred to as small resonator in the following. Its resonance frequencies and $Q$ factors are illustrated in \Figref{figdiameterQ}. The far field emission was characterized for angles between 160$^\circ$ and 290$^\circ$ at a resonance frequency of 170 GHz ($\lambda=1.76$~mm).

\begin{figure}[tbp]
\centering\includegraphics[width=\linewidth]{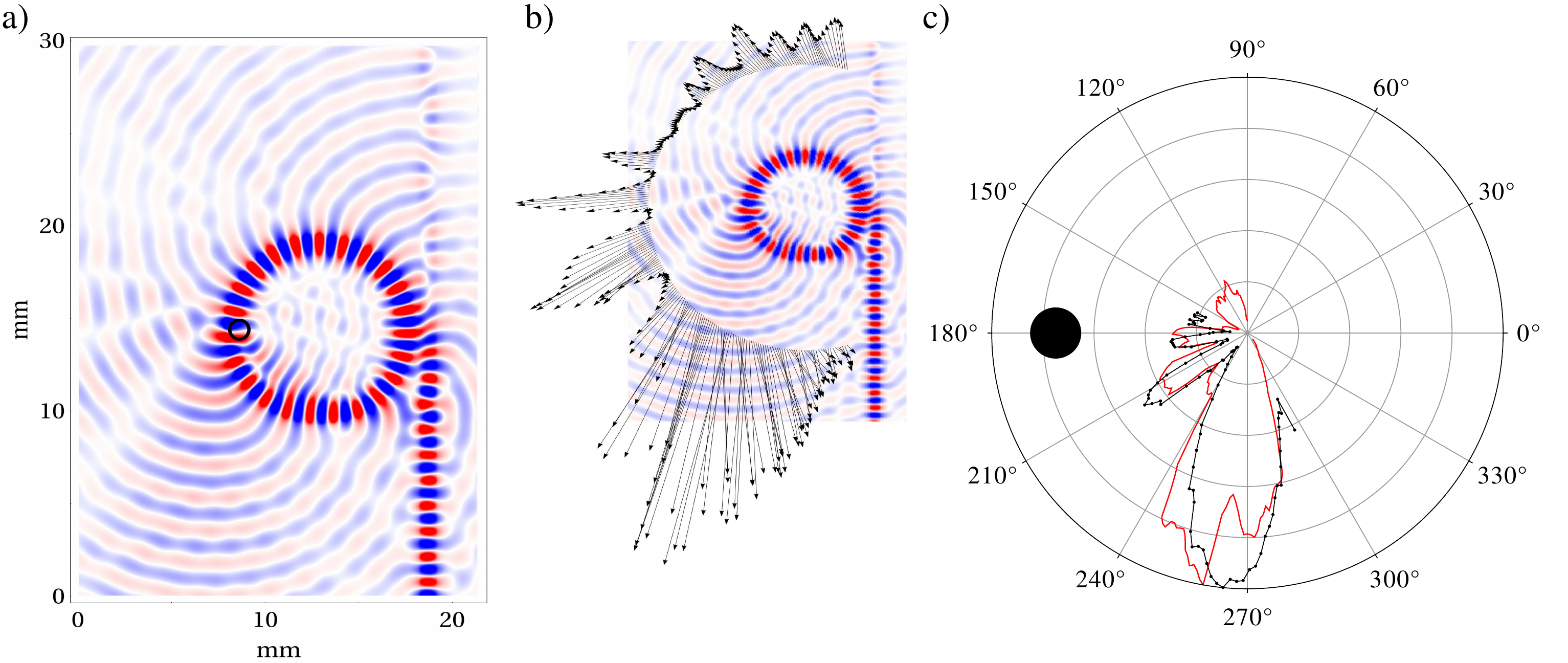}
\caption{Small Resonator ($R=5.2$ mm) with a hole at $180^\circ\pm5^\circ$. a) Snapshot of the FDTD calculation on a logarithmic scale. The small circle indicates the hole. b)Result of the Poynting vector analysis. The background shows a snapshot of the FDTD simulation. The arrows are the Poynting vectors, originating from the point at which they are evaluated, respectively. The Poynting vectors clearly reflect the pattern of the radiated field. The distance between the resonator edge and the circle on which the Poynting vectors are calculated is $\rho = 8$~mm.  c) Experimental data (black dots) as well as far-field radiation pattern predicted from the theoretical Poynting-analysis (red) on a linear scale. The dip in the theoretical emission intensity at $270^\circ$ is due to the waveguide region which was excluded from the Poynting vector analysis. In the experiment, the waveguide is curved and thus does not lie in this direction.
}\label{figSmallPoynting}
\end{figure}

%\begin{figure}[htbp]
%\includegraphics[width=12cm]{fig-small.pdf}
%\caption{Small Resonator ($R=5.2$ mm) with a hole at $180^\circ$. a) Snapshot of the FDTD calculation on a logarithmic scale. The small circle indicates the hole.  b) Experimental data (black dotted) as well as far-field radiation pattern predicted from the theoretical Poynting-analysis on a linear scale (red). The dip in the theoretical emission intensity at $270^\circ$ is due to the waveguide region which was excluded from the Poynting vector analysis. In the experiment, the waveguide is curved and thus does not lie in this direction. Overall, theory and experiment are in excellent agreement. }
%\label{figsmall}
%\end{figure}

The dimensions of our second resonator are $R=12.5$~mm, $r_h=1.25$~mm and $b=1$~mm. In the following this resonator is referred to as large resonator. The larger size was chosen to have less radiative losses of the unperturbed resonator and to simplify probing the near field with a probe waveguide.  The large resonator was studied at $\alpha_h=47^\circ$ and $\alpha_h=315^\circ$. 

Fig.~\ref{figSmallPoynting} a) shows a FDTD snapshot indicating the considered geometry. Fig.~\ref{figSmallPoynting} b) depicts the result of the Poynting vector analysis for the small resonator. The Poynting vector structure clearly resembles the interference pattern visible in the field configuration.  However, the radiation pattern cannot be read off directly from the Poynting vector structure, as the projection of the pattern onto the detection sphere has not yet been applied. The experimental data are compared to the prediction for the far-field pattern from the Poynting analysis in Fig.~\ref{figSmallPoynting} c).  The comparison between experiment and theory in (c) shows excellent agreement. The theoretical curve was obtained for  a detector size of $10^\circ$, and a detector distance of $10$~cm, consistent with the experiment. Apart from an overall scaling, no further free parameter had to be adjusted. The scaling was chosen such that the largest relative emission from the theoretical and the experimental data coincide. The main lobes as well as the intensity minima observed in the experiment are clearly reproduced in the theoretical analysis. The main difference arises at $270^\circ$, where the theoretical data exhibits a reduction in the emission intensity. This most likely is due to the waveguide, which in the theoretical calculation is situated at this emission angle, such that an angular range around $270^\circ$ had to be excluded from the Poynting vector analysis. In the experiment, the waveguide was bent, such that it did not interfere with the emitted radiation at this angle, see Fig~\ref{FigSetup}.
\begin{figure}[tbp]
\centering\includegraphics[width=\linewidth]{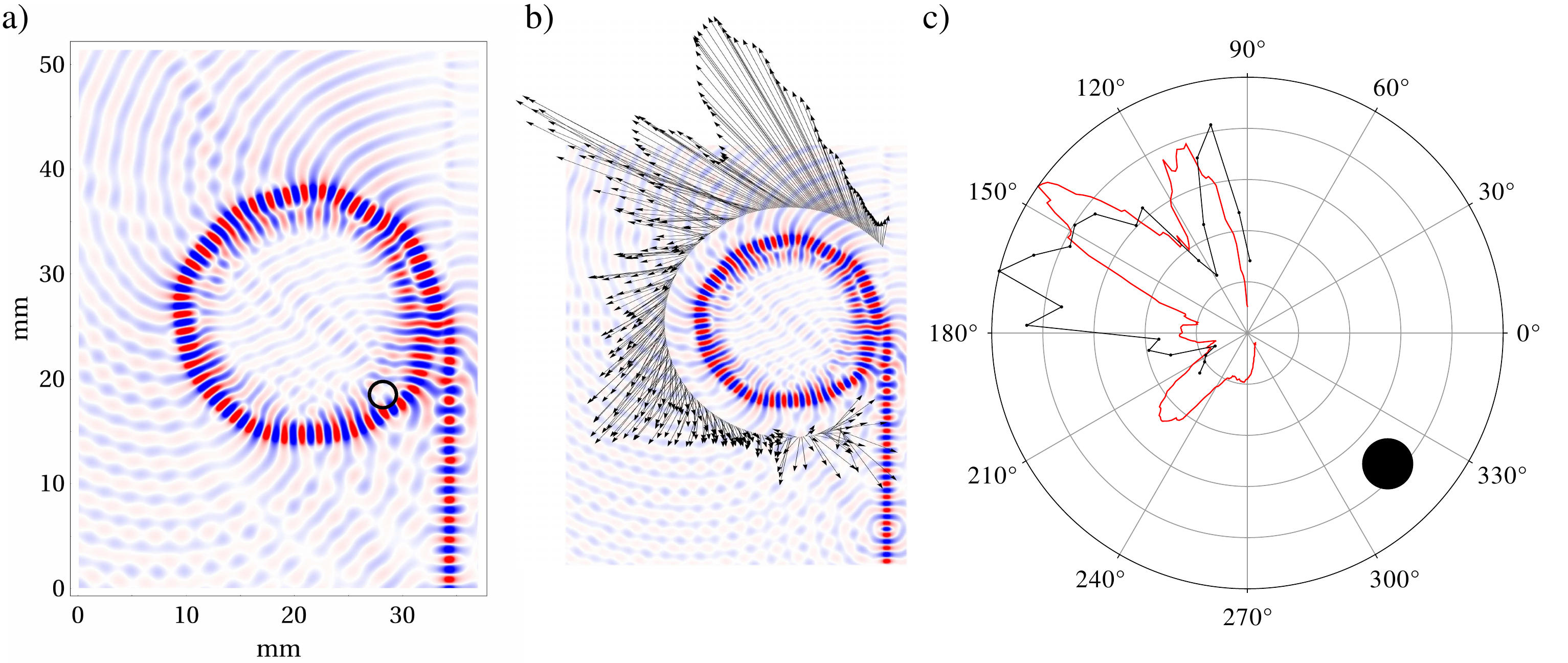}
\centering\includegraphics[width=\linewidth]{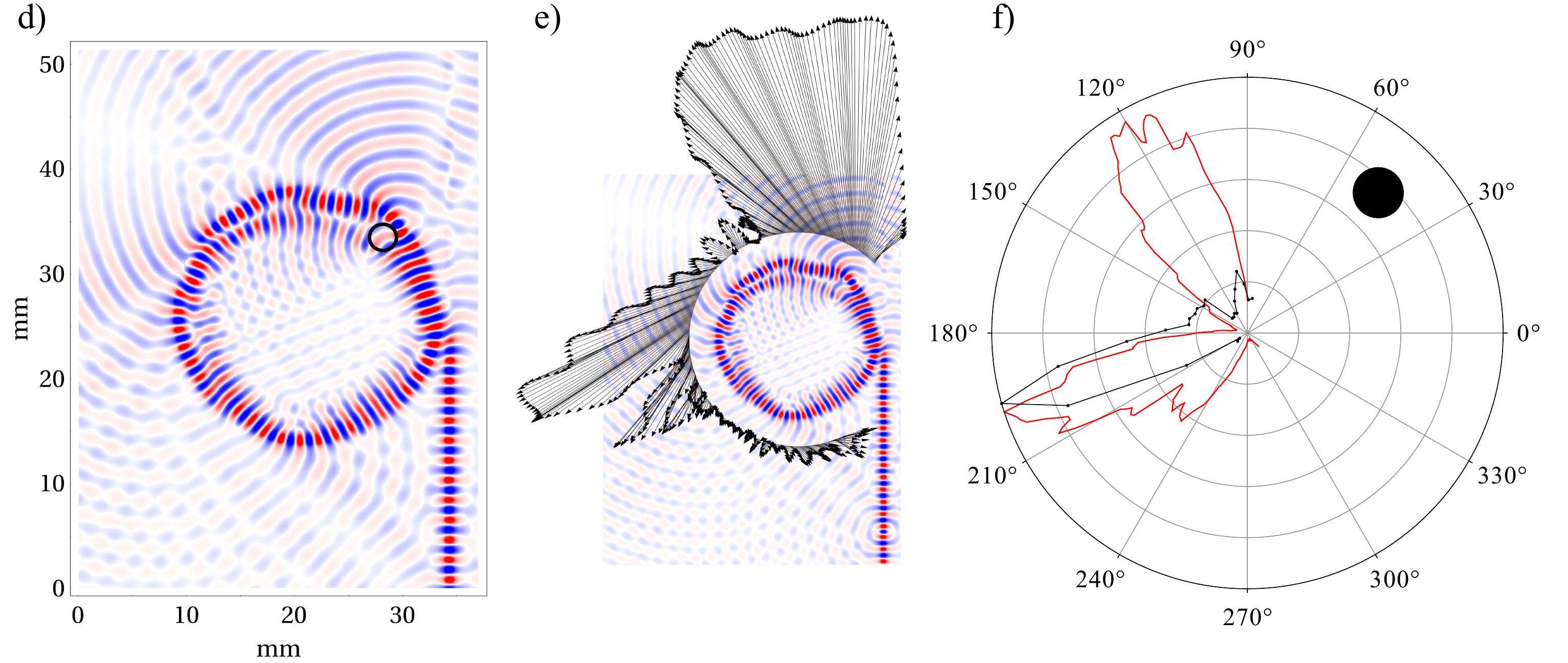}
\caption{Large Resonator, the hole with $d=12.5$ mm sits at $315^\circ\pm5^\circ$ (top row) and $47^\circ\pm5^\circ$ (bottom row); a) snapshot of the FDTD calculation. b) Result of the Poynting vector analysis, as in Fig.~\ref{figSmallPoynting}. The distance between the resonator edge and the circle on which the Poynting vectors are calculated is $\rho = 4$~mm. c) Experimental data (black dotted) as well as far-field radiation pattern predicted from the theoretical Poynting-analysis on a linear scale (red). While the second main lobe observed experimentally at $130^\circ-180^\circ$ is more narrow in the theoretical prediction, good overall agreement is achieved. d) snapshot of the FDTD calculation, the emission direction can be anticipated visually. e) Result of the Poynting vector analysis, as in Fig.~\ref{figSmallPoynting}.  The distance between the resonator edge and the circle on which the Poynting vectors are calculated is $\rho = 4$~mm. f) Experimental data (black dotted) as well as far-field radiation pattern predicted from the theoretical Poynting-analysis on a linear scale (red). While the first main lobe predicted from the theoretical calculation around $120^\circ$ is missing in the experimental data, experiment and theory agree well for the second lobe at $200^\circ$, most likely due to the slightly curved waveguide in the experiment.}\label{figlarge43Poynting} \label{figlarge223Poynting}
\end{figure}
The Poynting vector analysis and the comparison between experiment and theory for the large resonator with hole at $\alpha_h = 315^\circ$ are shown in Fig.~\ref{figlarge43Poynting}. Again, a detector size of $10^\circ$, and a detector distance of $10$~cm were chosen. Good qualitative matching between experiment and theory is achieved, even though the quantitative agreement is not as good as for the small resonator. The main difference is the width of the second lobe around $130^\circ-180^\circ$, which is wider in the experimental data than in the theoretical prediction. One possible origin for this could be slightly incorrect values for the index of refraction or the distance to the coupling waveguide to the resonator which could not be measured accurately, altering the perturbation of the resonator. Furthermore, the disk center is centered in the rotation stage. There may be some imaging error from radiation emitted at the disk boundary (being 12.5 mm off-axis) to the detector despite the 10\degree collimation angle (see Fig.~\ref{FigSetup}). This effect is much weaker for the small resonator, where theory and experiment agreed excellently.
%
%\begin{figure}[htbp]
%\centering\includegraphics[width=\linewidth]{fig-large-223-ba.pdf}
%\caption{Large Resonator, with hole of diameter $d=12.5$ mm at $47^\circ$. d) snapshot of the FDTD calculation, the emission direction can be anticipated visually. e) Result of the Poynting vector analysis, as in Fig.~\ref{figSmallPoynting}.  The distance between the resonator edge and the circle on which the Poynting vectors are calculated is $\rho = 4$~mm. f) Experimental data (black dotted) as well as far-field radiation pattern predicted from the theoretical Poynting-analysis on a linear scale (red). While the first main lobe predicted from the theoretical calculation around $120^\circ$ is missing in the experimental data, experiment and theory agree well for the second lobe at $200^\circ$. }\label{figlarge223Poynting}
%\end{figure}
%
Finally, the results for the large resonator with hole at $\alpha_h = 47^\circ$ are shown in Fig.~\ref{figlarge223Poynting}. The first main lobe at around $120^\circ$ predicted in the theoretical analysis is missing in the experimental data, but experiment and theory agree well for the second lobe around $200^\circ$. From the Poynting vector analysis, it can be concluded that the peak around $120^\circ$ originates from the coupling region of the resonator. 

Qualitatively, all emission patterns can be interpreted in the following way. On the one hand, there is emission in the direction of the tangent to the resonator at the hole position. This best visible for the small resonator around $270^\circ$ in Fig.~\ref{figSmallPoynting}, but also around $120^\circ$ in for the large resonator with hole at $\alpha_h = 47^\circ$, see Fig.~\ref{figlarge223Poynting} a-c). For the large resonator with $\alpha_h = 315^\circ$ in Fig.~\ref{figlarge43Poynting} d-f), this contribution is masked by the waveguide and the coupling region. A naive picture for this tangential emission is that a fraction of the light is deflected by the scatterer such that it acquires a larger incidence angle on the resonator surface compared to the unperturbed resonator mode, and therefore is able to scatter out of the resonator. 
Next to the main lobe in tangential direction, also side lobes with higher scattering angle are visible. These appear as side lobes at $210^\circ$ and $180^\circ$ in Fig.~\ref{figSmallPoynting} for the small resonator. Note that the structure at $180^\circ$ almost corresponds to a radial emission from the resonator, i.e., almost perpendicular to the energy flow in the resonator mode. The structure around $210^\circ$ corresponds to about $45^\circ$ deflection. From the simulation data, it also appears that similar higher deflection angles also occur towards the resonator center, even though they are masked in the emission pattern since they would have to pass both the resonator and the waveguide before they could reach the detector. 

Analogous contributions due to larger scattering angles are also visible for the large resonator. For example, for $\alpha_h = 315^\circ$ in Fig.~\ref{figlarge43Poynting}, the structure at $110^\circ$ corresponds to approximately $45^\circ$ deflection, and the one at $150^\circ$ to approximately $90^\circ$ deflection. Finally, the large resonator with $\alpha_h=47^\circ$ also exhibits a contribution with deflection angle around $90^\circ$, which is observed around $200^\circ$ emission direction. Interestingly, in this case the corresponding emission in $45^\circ$ deflection direction is not observed. One reason for this could be interference with light scattered from the coupling region between resonator and waveguide. 

The emission patterns for the two hole positions shown in Fig.~\ref{figlarge43Poynting} with respect to the coupling position substantially differ. This is in agreement with our findings in Fig.~\ref{fig:2Dimage}. The waveguide already perturbs the modal structure. This effect interferes with the effect of the hole. Both perturbations have to be taken into account in order to determine the emission pattern.

\section{Conclusion}
In conclusion, we demonstrated both experimentally by measurements in the THz domain and theoretically by a finite difference time domain simulation, that a finite scatterer close to the circumference of the whispering gallery mode resonator allows for directional outcoupling without ruining the quality factor. The main emission direction was tangential, originating at the scatterer. However, there exist side lobes with larger emission angles. Influence of other perturbations such as a coupling waveguide have to be taken into account in order to numerically determine the emission direction. With a finite difference time domain method, we achieved good agreement between the experimental data and the theoretical modeling in all investigated cases.

\section*{Acknowledgement}
We thank L.\ Hong and A.\ C.\ Gossard at MRL Santa Barbara, CA, USA, for growing the substrate of the THz devices. We further would like to thank Gerd Leuchs and his devision at the MPL for the creative atmosphere and the financial support.

\bibliography{directionalTHz}
\bibliographystyle{plainnat}

\end{document}